\documentclass[journal]{IEEEtran}
\usepackage[pdftex]{graphicx}
\usepackage{arydshln}
\usepackage[justification=centering]{caption}
\usepackage{amsmath}
\usepackage{nccmath}
\usepackage{xcolor}
\definecolor{myblue}{RGB}{84,36,125}
\definecolor{mybrown}{RGB}{115,31,25}
\usepackage{color,soul}
\usepackage{float}

\title{S-vectors and TESA: Speaker Embeddings\\ and a Speaker Authenticator Based on \\ Transformer Encoder}

\author{N J Metilda Sagaya Mary, S Umesh, Sandesh V Katta
\thanks{N~J~Metilda~Sagaya~Mary and S~Umesh are with the Department
of Electrical Engineering, Indian Institute of Technology Madras, Chennai 600036, India email:(metty789@gmail.com, umeshs@ee.iitm.ac.in)}
\thanks{Sandesh V Katta was with the Department
of Electrical Engineering, Indian Institute of Technology Madras, Chennai 600036, India email:(sandesh.katta97@gmail.com)}}

\begin{document}
\begin{minipage}[t]{0.75\paperwidth}
\vspace{2cm}
{\Huge IEEE Copyright Notice}\\

© 2021 IEEE. Personal use of this material is permitted. Permission from IEEE must be obtained for all other uses, in any current or future media, including  reprinting/republishing this material for advertising or promotional purposes, creating new collective works, for resale or redistribution to servers or lists, or reuse of any copyrighted component of this work in other works

\vspace{2cm}
{\Large
\textbf{Accepted to be Published in: IEEE Transactions on Audio, Speech and Language Processing. }}
\end{minipage}
\pagebreak 

\maketitle

\begin{abstract}
One of the most popular speaker embeddings is x-vectors, which are obtained from an architecture that gradually builds a larger temporal context with layers. In this paper, we propose to derive speaker embeddings from Transformer's encoder trained for speaker classification. Self-attention, on which Transformer's encoder is built, attends to all the features over the entire utterance and might be more suitable in capturing the speaker characteristics in an utterance. We refer to the speaker embeddings obtained from the proposed speaker classification model as s-vectors to emphasize that they are obtained from an architecture that heavily relies on self-attention. Through experiments, we demonstrate that s-vectors perform better than x-vectors. In addition to the s-vectors, we also propose a new architecture based on Transformer's encoder for speaker verification as a replacement for speaker verification based on conventional probabilistic linear discriminant analysis (PLDA). This architecture is inspired by the next sentence prediction task of bidirectional encoder representations from Transformers (BERT), and we feed the s-vectors of two utterances to verify whether they belong to the same speaker. We name this architecture the Transformer encoder speaker authenticator (TESA). Our experiments show that the performance of s-vectors with TESA is better than s-vectors with conventional PLDA-based speaker verification.

\end{abstract}


\section{Introduction}
\IEEEPARstart{S}{peaker} verification uses speech as a biometric to verify the identity claimed by the speaker. There are two types of speaker verification systems: text-dependent and text-independent. Text-independent systems are flexible, as there is no constraint on the text spoken by the speaker. Most of the research in this area is focused on obtaining a single fixed-dimension vector representing an utterance. These vectors are then scored to verify the speaker's identity and are termed speaker embeddings. Any speaker embedding should enhance interspeaker variability and suppress intraspeaker variability while scoring.

One of the earliest speaker embeddings is i-vectors \cite{i-vec} extracted from the universal background model-Gaussian mixture model. With the increase in data available for training, speaker embeddings based on deep learning methods have gained in popularity.

A common approach in all the deep learning methods addressing the speaker verification task is to perform speaker classification first \cite{Snyder2017, SNYDER-18, CNN2, CNN1, spk-ver-w}. Then, utterance-specific fixed-dimension embeddings are obtained from the speaker classification network by different pooling methods. These embeddings are then fed to a speaker verification system to validate the identity claimed by the speaker. One such embedding is x-vectors \cite{SNYDER-18} extracted from a speaker classification architecture based on time-delay neural network (TDNN) \cite{TDNN} with pooling. Hereafter, the term x-vectors refers to the original TDNN-based x-vectors in \cite{SNYDER-18}. The x-vectors along with speaker verification based on probabilistic linear discriminant analysis (PLDA) \cite{PLDA} have significantly outperformed i-vectors.

Recently, speaker embeddings extracted from speaker classification architectures based on  residual networks (ResNets) \cite{resnet} \cite{CNN2,CNN1, spk-ver-w,TF,TFC} with pooling, which consider spectrogram representations of speech as images, have outperformed x-vectors. It should be noted that ResNets are pretrained models and are pretrained with a considerable number of images. These pretrained models are then retrained with speech data to perform speaker classification with standard softmax or additive margin softmax loss \cite{add-m}. Then, speaker verification is performed with loss functions such as contrastive \cite{contra1, contra2} or relation loss \cite{spk-ver-w}.

Different strategies called pooling methods are employed in speaker classification networks to obtain fixed-dimension embeddings from variable length utterances, as mentioned before. These pooling methods can be classified into nontrainable and trainable approaches. Average \cite{CNN2} and statistics pooling \cite{SNYDER-18} are nontrainable pooling methods. In average pooling, all the frames of the given dimension $ D \times 1 $ are averaged to obtain a single $ D \times 1$ dimension vector. In statistics pooling, both mean and standard deviation vectors are concatenated to obtain a $2D \times1$ vector. Attentive statistics \cite{Okabe}, self-attention \cite{Lin2017,Zhu2018}, 2D self-attention \cite{sun2019speaker}, NetVLAD \cite{netvlad}, and GhostVLAD \cite{ghostvlad} are trainable pooling methods. Not all frames are important in identifying a speaker. Therefore, attentive statistics and self-attention pooling aim to assign weights to the frames while calculating utterance level statistics. 2D self-attention pooling tries to incorporate an additional self-attention layer to pool the embeddings from different models. NetVLAD learns a $K \times D_{k}$ matrix, where $K$ represents the number of clusters taken and $D_{k}$ is the dimension of clusters. This matrix is then flattened and fed for speaker classification after some processing. GhostVLAD is similar to NetVLAD, with a few clusters being dropped.

Recently, different attention mechanisms \cite{Att1,Att2} have been explored in natural language processing (NLP) and speech processing areas. A new architecture named Transformer \cite{Transformer} has outperformed many existing NLP models. Transformers have been successfully applied in both automatic speech recognition and text-to-speech tasks \cite{sptrn,li2018neural,LR,CS}. A Transformer is built on self-attention across time. Some architectures for speaker verification have explored incorporating attention at a model level in frameworks different from Transformer. In \cite{TF}, attention in both time and frequency was explored in a ResNet framework. In a TDNN framework, channelwise attention has shown improvement in speaker verification performance \cite{desplanques2020ecapa}. In \cite{TFC}, combining time-frequency attention with channel attention hierarchically at the model level was explored in a ResNet framework.

A Transformer has two modules: an encoder and a decoder. These two modules are built on self-attention and interact through source-attention. In the Transformer's encoder, information from all the frames is accounted for by the self-attention networks in every layer. Self-attention is built on dot products. As dot products are a kind of similarity measure that might better capture speaker characteristics in an utterance, we were motivated to explore embedding extraction from Transformer's encoder trained for the speaker classification task. It should be noted that this idea is different from attention-based pooling \cite{Okabe,Lin2017, Zhu2018, sun2019speaker}, where attention is employed in the pooling only. At the time of submission, we also noticed a work that briefly discusses the extraction of embeddings from Transformer encoder \cite{parallel}. \cite{safari2020selfattention} also explored the extraction of speaker embeddings from a Transformer encoder but with the objective of reducing the parameter count.

Different self-supervised pretrained models based on Transformers have given state-of-the-art performance in various NLP tasks. One such model is bidirectional encoder representations from Transformers (BERT) \cite{devlin-etal-2019-bert}. BERT is a Transformer encoder with some special input tokens and is trained on two tasks simultaneously: masked language modeling and next sentence prediction. Reconstruction of masked speech frames such as the masked words being predicted in the masked language modeling task of BERT was explored in \cite{liu2020tera}, and the features extracted from it performed well in speaker classification as a downstream task. The next sentence prediction task of BERT takes two sentences as input, and the model is trained to predict whether the second sentence follows the first sentence. This next sentence prediction task motivated us to propose Transformer encoder speaker authenticator (TESA) in place of conventional PLDA.

Other works in speaker verification have been in the area of using different features. The impact of mel frequency cepstral coefficients (MFCCs) of different dimensions and fbank features in the x-vector framework was explored in \cite{Lin2020}. Raw speech with a convolutional neural network-based speaker embedding extractor has also given the improvement in \cite{Lin2020}.

Our main contributions in this paper are as follows:
\begin{itemize}
    \item A Transformer encoder-based speaker classification architecture is proposed, from which we obtain better speaker embeddings. We call the embeddings obtained from this architecture s-vectors. We replaced the TDNN part of the x-vector model's architecture with Transformer's encoder to obtain the s-vector model's architecture without changing the input MFCC dimension.
    \item A Transformer encoder-based architecture for speaker verification is proposed, where we feed speaker embeddings from two utterances together, to verify whether they belong to the same speaker. We call this architecture the Transformer encoder speaker authenticator (TESA). TESA is inspired by the next sentence prediction task of BERT.
    \item Through experiments, we demonstrate that both the s-vectors and TESA outperform their existing baseline counterparts x-vectors and PLDA on both VoxCeleb-1 and VoxCeleb-2 data.
\end{itemize}

\section{X-vectors}
\subsection{Architecture}
\begin{figure}[t]
\centering
\includegraphics[width=0.8\linewidth]{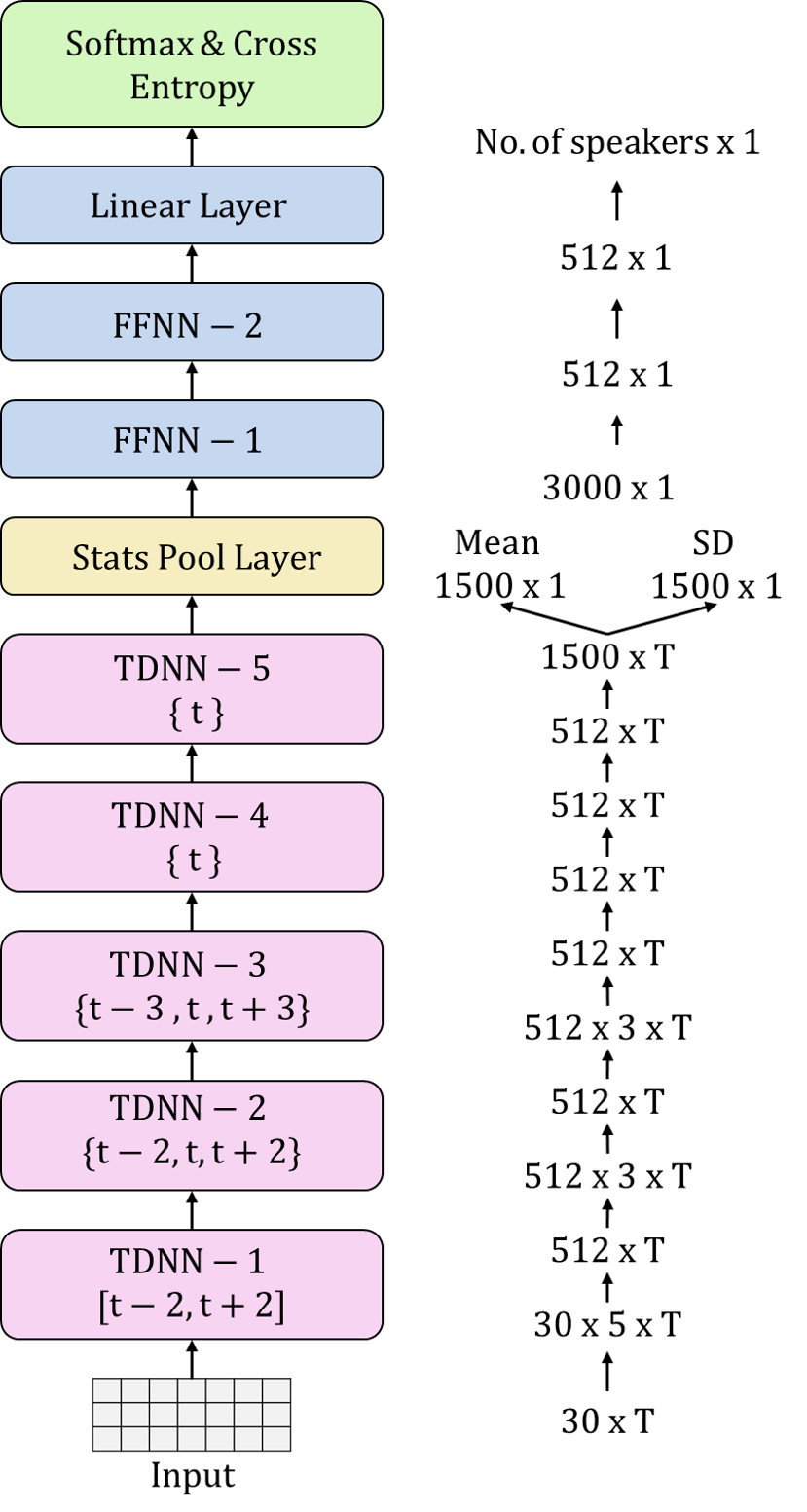}  \caption{X-vectors architecture}
\label{fig:x-vector}
\end{figure}
 The architecture of the x-vector model is shown in Fig.~\ref{fig:x-vector}. We used 30-dimensional MFCC features instead of the 24-dimensional MFCC features in \cite{SNYDER-18}. In the first TDNN layer (TDNN-1), at any time step $t$, \{$t-2, t-1, t, t+1, t+2$\} frames are spliced and presented as input. TDNN-2 takes \{$t-2, t, t+2$\}, and TDNN-3 takes \{$t-3, t, t+3$\} spliced frames of the previous layer as input. TDNN-4 and TDNN-5 take just the $t\textsuperscript{th}$ frame as input. To aggregate the statistics over the entire utterance, this system uses a statistics pooling layer (stats pool layer). This layer computes the mean and standard deviation across each dimension over the entire utterance, resulting in a single \textit{3,000} $\times$ \textit{1} vector. This single vector is representative of the entire utterance. Then, this vector is passed through two feedforward neural networks (FFNNs) and then to the output layer with cross-entropy as the criterion for speaker classification. All nonlinearities used are rectified linear units (ReLUs), and batch normalization is performed at every stage. The embeddings extracted from FFNN-1 before nonlinearity are termed x-vectors. This model is the baseline with which we compare our proposed s-vector model.

\subsection{PLDA-based Speaker Verification}
The x-vectors method uses PLDA to generate scores for the trial pairs. If the scores are higher than the set threshold, utterances in the pair belong to the same speaker and vice versa. Linear discriminant analysis helps verify seen speakers and cannot handle the unseen speaker scenario optimally. Therefore, PLDA attempts to fit a Gaussian mixture model with a continuous class variable to generalize for unseen speakers.

Let the probability of generating the data samples $x$ from a given class $y$ be given by

\begin{equation}\label{eq:1}
   P(x \vert y) \sim \mathcal{N}(x \vert y,\phi_{w}), \
\end{equation}

\noindent where $\phi_{w}$ is the within-class covariance matrix and is common for all classes. To enable efficient handling of unseen classes, a Gaussian class prior is taken as

\begin{equation}\label{eq:2}
    P(y) = \mathcal{N}(y \vert m,\phi_{b}), \\
 \end{equation}

\noindent where $\phi_{b}$ is the between-class covariance matrix and $m$ is the mean of all data samples. $\phi_{w}$ is positive definite, and $\phi_{b}$ is positive semidefinite. A matrix $V$ simultaneously diagonalizes both $\phi_{w}$ and $\phi_{b}$ as given by

\begin{equation}\label{eq:3}
    \begin{split}
        V^{T} \phi_{b} V & = \psi \\
        V^{T} \phi_{w} V & = I, \\
    \end{split}
\end{equation}

 \noindent where $\psi$ is a diagonal matrix and $I$ is the identity matrix. The class variable $v$ and an example of a class $u$ in the latent space are given by
\begin{equation}\label{eq:4}
\begin{split} 
    v & \sim \mathcal{N}(. \vert 0,\psi) \\
    u & \sim \mathcal{N}(. \vert v, I). \\
\end{split}
\end{equation}

\noindent $u$ and $v$ are related to the feature space by
\begin{equation}\label{eq:5}
\begin{split} 
    y & = m + Av \\
    x & = m + Au, \\
\end{split}
\end{equation}
\noindent where $A$ is $V^{- T}$.

Now, we have to test whether two examples (trial pair) from unseen classes belong to the same class. In the speaker verification task, classes are speakers, and examples are the embeddings (x-vectors or proposed s-vectors) extracted from the utterances. To test whether the embeddings $u_{1}$ and $u_{2}$ of the two example utterances belong to the same speaker, the logarithm of the ratio of the likelihoods $P(u_{1},u_{2})$ and $P(u_{1})$ $P(u_{2})$ is computed. This is the PLDA score for the given trial pair, and if the score is greater than the set threshold, the utterances are taken to be from the same speaker and vice versa.

\section{Proposed S-vectors}
Existing speaker classification architectures gradually build temporal context with layers\textbf{,} as mentioned before. To capture the speaker characteristics better, we used the Transformer encoder, which is based on self-attention in our architecture. Its strength is that it is not restricted to finite context and attends to all frames in each of its layers as opposed to \cite{Lin2017, Okabe, Zhu2018, sun2019speaker}, where attention is employed only during pooling.
\begin{figure}[t]
  \centering
  \includegraphics[width=0.8\linewidth]{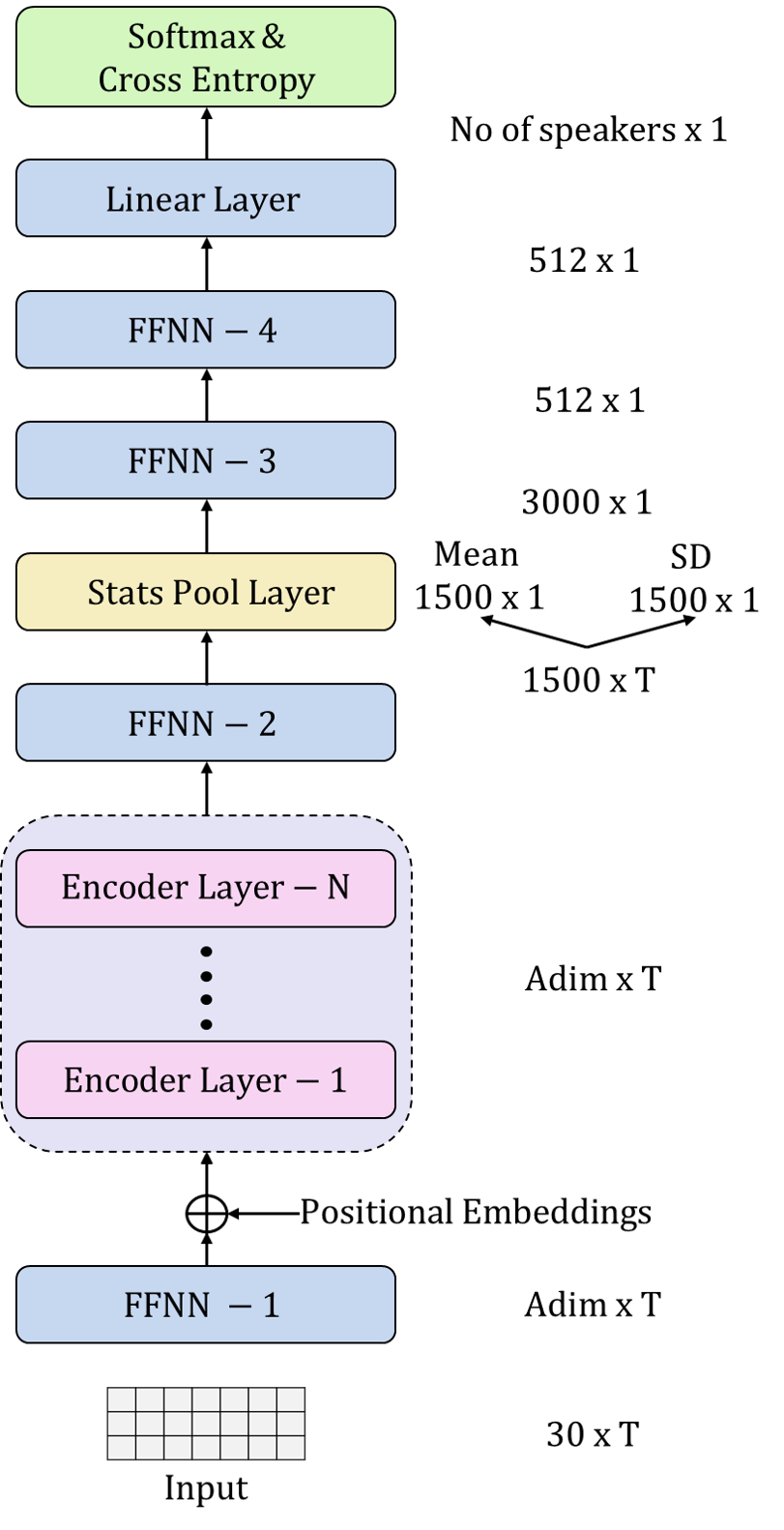}
  \caption{S-vectors architecture}
  \label{fig:Sav}
\end{figure}

\subsection{Architecture}
To derive s-vectors, we replaced the TDNN in the baseline x-vector model with the encoder of the Transformer \cite{Transformer}, as shown in Fig.~\ref{fig:Sav}. Input is the same \textit{30} $ \times $ \textit{T} dimension MFCC features. We used the same training utterances as used by the x-vector system to ensure a fair comparison. \textit{30} $ \times $ \textit{T}-dimensional MFCC features are transformed into the \textit{attention dimension (Adim)} $ \times $ \textit{T} by FFNN-1 and fed to the above encoder layers after adding the position embeddings. Multihead self-attention is performed at every encoder layer. The encoder layer is explained in detail in the next section. The resultant \textit{Adim} $ \times$ \textit{T} final encoder layer's output is then taken to \textit{1,500} $ \times $ \textit{T} through FFNN-2. Statistics pooling on these vectors results in a single \textit{3,000} $ \times $ \textit{1} vector (\textit{1,500} $ \times $ \textit{1} mean and \textit{1,500} $ \times $ \textit{1} standard deviation). This \textit{3,000} $ \times $ \textit{1} vector is then taken to \textit{512} $ \times $ \textit{1} and then to \textit{512} $ \times $ \textit{1} again by two FFNNs. The resultant vector is then presented to a classification layer. In all FFNNs except FFNN-2, we used ReLU. We used leaky ReLU (negative slope = 0.01) in FFNN-2 to stabilize the gradients flowing through the standard deviation of the stats pool layer. Speaker embeddings are extracted from the affine part of FFNN-3. We call the speaker embeddings extracted from our proposed model s-vectors.

\begin{figure}[t]
  \centering
  \includegraphics[width=0.4\linewidth]{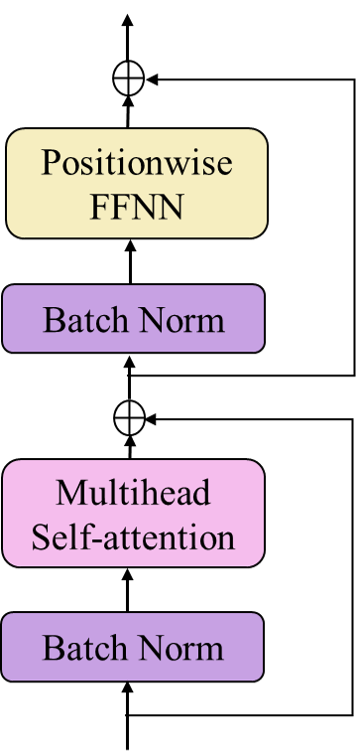}
  \caption{Encoder layer architecture}
  \label{fig:Internals of Encoder Layer}
\end{figure}
\subsubsection{Encoder Layer}
Each encoder layer is made of a multihead self-attention network and a positionwise FFNN, as shown in Fig.~\ref{fig:Internals of Encoder Layer}. Positionwise FFNN is composed of two FFNNs. The first FFNN converts the input \textit{Adim} $ \times$ \textit{T} to \textit{encoder units} $ \times$ \textit{T}, and the second FFNN projects the output of the first FFNN back to \textit{Adim} $ \times$ \textit{T}. Batch normalization is performed after adding residuals in every stage of the encoder layer. Batch normalization gave better performance than the usual layer normalization in the Transformers.

In the multihead self-attention network of the Transformer encoder, to obtain the output of a head $i$ ($O^{i}$), the input (U) is converted to queries ($Q^i$), keys ($K^i$) and values ($V^i$) through the respective $W_{Q}^i, W_{K}^i, W_{V}^i$ matrices as:

\begin{gather}\label{eq.6}
    U \in   R^{ T \times Adim } \nonumber\\ 
      W_Q^i , W_K^i \in   R^{ Adim \times d_k} \nonumber\\
      W_V^i \in R^{Adim \times d_v} \nonumber \\
    Q^i = U \times W_Q^i  \\
    K^i = U \times W_K^i  \nonumber\\
    V^i = U \times W_V^i.  \nonumber
\end{gather}

 \noindent $O^{i} \in R^{T\times d_v}$ is the weighted average of the frames of $V^{i}$. The weights of the frames of $V^{i}$ are obtained by,

\begin{equation}\label{eq.7}
    O^{i} =  \frac{softmax(Q^iK^{i^T}) }{\sqrt{d_k}}  V^i ,
\end{equation}

\noindent where softmax is computed along the row. In the case of multihead self-attention with $P$ heads, the final resultant output ($O$) is given by

\begin{equation}\label{eq.8}
  O  = Concat(O^1,O^2,...,O^p) \in R^{T \times (d_v\times P)}.
\end{equation}

\noindent The value of $d_k$ and $d_v$ is set to $\frac{Adim}{P}$.

\subsection{Datasets}
We used two datasets in training our models: VoxCeleb-1 \cite{vox1} and VoxCeleb-2 \cite{spk-ver-w}. VoxCeleb-1 consists of recordings from 1,251 speakers and over 0.1 million utterances extracted from celebrity interview videos on YouTube. VoxCeleb-2 is approximately ten times larger than VoxCeleb-1. It consists of recordings from 6,112 speakers and has approximately a million utterances extracted from celebrity interview videos on YouTube. The minimum length of utterances in VoxCeleb-1 and VoxCeleb-2 is 3 seconds, and the maximum length is 20 seconds. The two datasets together amount to approximately 2,000 hours, and the male-female gender ratio is 61\%-39\%. The utterances are from speakers around the world belonging to different professions.
Unless specified, all the models in this paper are trained on the whole of VoxCeleb-2 along with the Dev set of VoxCeleb-1. This training set is referred to as VoxCeleb-1+2 in this paper. Models are evaluated on VoxCeleb-1 Test. The VoxCeleb-1 Test consists of 4,874 utterances from 40 speakers and has 37,720 trial pairs.

\subsection{Training Details}
We used Kaldi's \cite{kaldi} VoxCeleb v2 recipe, which is the implementation of x-vectors in \cite{SNYDER-18} with 30-dimensional MFCC features, to obtain our baseline. The same recipe was used to extract features and prepare data for s-vectors. After feature extraction, energy-based voice activity detection (VAD) is performed. The training data are then augmented with reverberation and noise in the same way as Snyder et al. for their x-vector model using VoxCeleb-1+2 data \cite{SNYDER-18}. Reverberation examples are taken from the RIRS database \cite{rirs}, and noise examples are taken from the MUSAN database \cite{musan}. Cepstral mean normalization (CMN) is applied to the augmented utterances, and nonspeech frames are removed. They are then chunked to generate training examples, similar to the baseline x-vector model.

Our s-vector model was trained in ESPnet \cite{esp}. After data augmentation with RIRS and MUSAN, Kaldi's VoxCeleb v2 recipe chunks the data into random lengths with different start frames to generate training examples and writes it as .ark files before training. These ark files are not compatible with ESPnet. Therefore, we obtained chunk information, such as the start frame and chunk length, and performed chunking in ESPnet before passing an utterance for training. Therefore, the s-vector models were trained on the same data as that of the baseline x-vector model for a fair comparison. Other details of the s-vector training conducted in ESPnet are presented in Table~\ref{tab:a}.

\begin{table}
\renewcommand{\arraystretch}{1.3}
\caption{T\textsc{raining} D\textsc{etails} \textsc{of} S\textsc{-vector} M\textsc{odels} T\textsc{rained on} V\textsc{oxCeleb-1+2}}
 \label{tab:a}
 \centering
\begin{tabular}{c|c}
\hline
Parameter &VoxCeleb-1+2 \\\hline \hline
Position Encoding  &Sinusoidal \\
Encoder Units      &2,048    \\
Normalize Before   &True   \\
Learning Rate      &10    \\ 
Batch Size         &$\approx$ 100\\
Optimizer          &Noam   \\
Dropout             &0.1 \\
Gradient-clip       &5 \\
Warm-up Steps      &25,000      \\ 
Normalization      &Batch Norm \\ \hline
\end{tabular}
\end{table}

\begin{figure}
  \centering
  \includegraphics[width=0.7\linewidth]{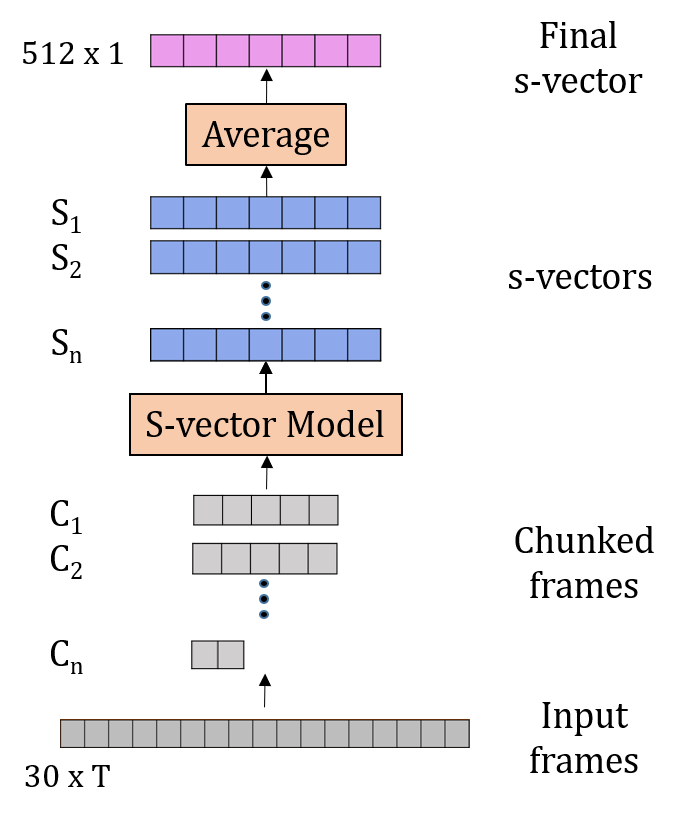}
  \caption{Extraction of s-vectors by chunking}
  \label{fig:chunking}
\end{figure}
\section{Results with PLDA and Discussion}
In this section, we discuss the results of the s-vector model with conventional PLDA scoring. We utilized Kaldi's VoxCeleb v2 recipe for PLDA scoring. Before feeding the utterances for s-vector extraction, CMN was performed, and nonspeech frames were removed similar to the x-vectors using an energy-based VAD system. Then, the resultant utterance frames were chunked, with each chunk being 300 frames, and the remaining frames were taken as another chunk. There was no overlap between the chunks. Embeddings were then extracted for each chunk and then averaged to obtain the s-vector, as shown in Fig~\ref{fig:chunking}. We tried different LDA dimensions and found that the optimal dimension was 250 for the VoxCeleb-1+2 dataset.

\subsection{Evaluation Metrics}
The standard equal error rate (EER) and detection cost function (DCF) were used as the evaluation metrics to compare the baseline x-vector and our proposed s-vector model. EER refers to the value at which false alarm and miss error rates become equal. DCF is a weighted linear combination of false alarm and miss error rates. For DCF calculation, we assume P\textsubscript{target} = 0.01 (or 0.001), while C\textsubscript{miss} = 1 and C\textsubscript{false alarm} = 1.

\begin{table}
\renewcommand{\arraystretch}{1.3}
 \caption{V\textsc{oxCeleb-1+2} R\textsc{esults on} V\textsc{oxCeleb-1} T\textsc{rials}}
 \label{tab:vox2}
 \centering
 \begin{tabular}{c|c|c|c}
 \hline
Model & \% EER & DCF (0.01) &DCF (0.001) \\ \hline
\hline
x-vector baseline &3.05 &0.33 &0.5\\
3L-256D-4H-S &3.35 &0.34 &0.53\\
6L-256D-4H-S &2.87 &0.31 &0.51\\
\textbf{6L-512D-8H-S} &\textbf{2.67} &\textbf{0.30} &\textbf{0.44}\\
9L-512D-8H-S &2.72 &0.29 &0.53\\ \hline
\multicolumn{4}{p{7.0cm}}{Note: 6L-512D-8H-S denotes a 6 encoder layer (L), 512 Adim (D), and 8 head (H) s-vector (S) model.}
\end{tabular}
\end{table}
\begin{figure}
\centering
  \includegraphics[width=0.9\linewidth]{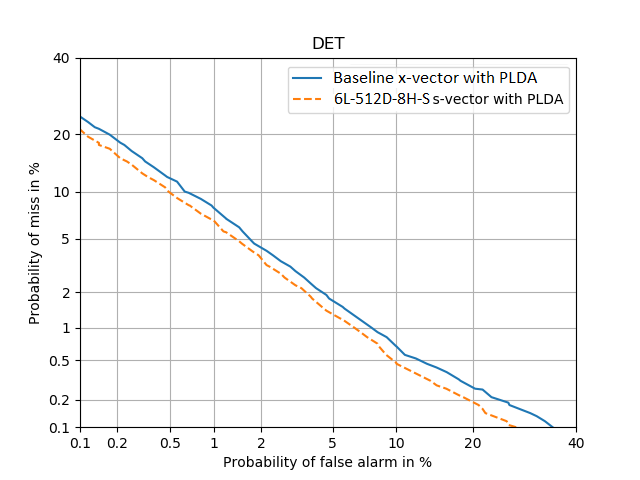}
  \caption{Comparison of DET curves of the 6L-512D-8H-S model with the baseline x-vector model when trained on VoxCeleb-1+2 data }
  \label{fig:DET_vox2}
\end{figure}

\subsection{Finding the Optimal Hyperparameters for the S-vector Model when Trained on VoxCeleb-1+2 Data}
We analyzed the proposed s-vector architecture by varying the hyperparameters for VoxCeleb-1+2 data. \% EER and DCF for the different numbers of layers and other hyperparameters taken are presented in Table~\ref{tab:vox2}. 6L-512D-8H-S in Table~\ref{tab:vox2} denotes a 6 encoder layer (L), 512 Adim (D), and 8 head (H) s-vector (S) model. The detection error tradeoff (DET) curves for the best performing s-vector and the baseline x-vector model are presented in Fig.~\ref{fig:DET_vox2}. We see that the s-vector model outperforms the baseline x-vector model in terms of \% EER by 12.5\% relative, and DCF values are also better. The number of parameters in the 6L-512D-8H-S model is 25.3 million, and that of the baseline x-vector model is 4.3 million. Even the 6L-256D-4H-S model surpasses the x-vector baseline except for DCF (0.001) and has 13.8 million parameters.

\vspace{-1cm}

\subsection{Finding the Optimal Hyperparameters for the S-vector Model when Trained on VoxCeleb-1 Data}
\begin{figure}
  \centering
  \includegraphics[width=0.9\linewidth]{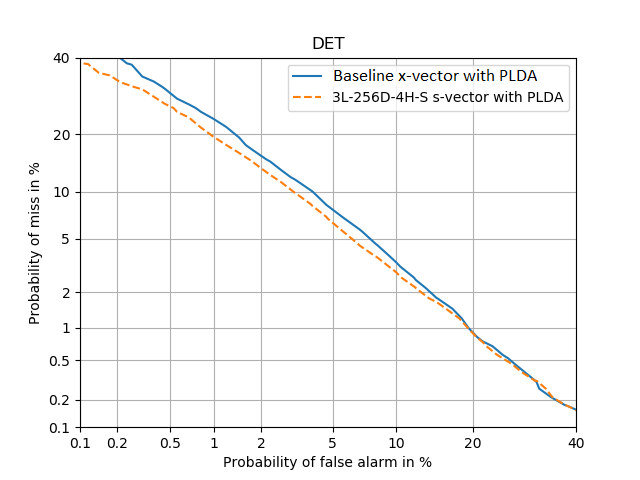}
  \caption{Comparison of DET curves of 3L-256D-4H-S model with the baseline x-vector model when trained on VoxCeleb-1 data}
  \label{fig:DET_vox1}
\end{figure}

 In this section, we show that the s-vector model outperforms the baseline x-vector model even with smaller training data. Both s-vector and the baseline x-vector models were trained with VoxCeleb-1 Dev data. These data are almost ten times smaller than the VoxCeleb-1+2 data. \% EER and DCF for the different numbers of layers taken for the s-vector model are presented in Table~\ref{tab:b}. We see that all three s-vector models consistently perform better than the baseline x-vector model in terms of \% EER. The 3L-256D-4H-S model outperforms the baseline x-vector model trained on the same VoxCeleb-1 Dev data in both \% EER and DCF. Therefore, the s-vectors outperform the x-vectors even in a dataset much smaller than VoxCeleb-1+2. The DET curves for the x-vector baseline and the best performing 3L-256D-4H-S model, when trained on the smaller VoxCeleb-1 Dev data, are presented in Fig.~\ref{fig:DET_vox1}.
\begin{table}
\renewcommand{\arraystretch}{1.3}
 \caption{V\textsc{oxCeleb-1} R\textsc{esults on} V\textsc{oxCeleb-1} T\textsc{rials}}
 \label{tab:b}
 \centering
 \begin{tabular}{c|c|c|c} 
 \hline
Model & \% EER & DCF (0.01) &DCF (0.001) \\ \hline
\hline
x-vector baseline &6.26  &0.54    &0.64 \\
2L-256D-4H-S     &5.80    &0.48     &0.65\\ 
\textbf{3L-256D-4H-S}   &\textbf{5.63}     &\textbf{0.46}    &\textbf{0.62}\\ 
4L-256D-4H-S      &5.50    &0.50    &0.64 \\\hline
\multicolumn{4}{p{7.0cm}}{Note: 3L-256D-4H-S means it is a 3 encoder layer (L), 256 Adim (D), and 4 head (H) s-vector (S) model.}
\end{tabular}
\end{table}

\subsection{Effect of Chunking}
In this section, we analyze the performance for different chunk lengths. We chose to chunk the utterance and obtain the embeddings for each utterance because of the position embeddings in our architecture. We expected that feeding whole utterances would lead to unseen positions in the input and result in poor speaker embeddings. Embeddings extracted for each chunk in an utterance are averaged to obtain the final embedding. The performance of the 6L-512D-8H-S model for different chunk lengths is presented in Table~\ref{tab:gg}. We see that the performance is better for chunk lengths of 300 and 500. A chunk length of 300 gives a good DCF (0.001) value, and we take it as the optimal chunk length.

\begin{table}
\renewcommand{\arraystretch}{1.3}
\caption{E\textsc{ffect of} C\textsc{hunking on} 6\textsc{-layer} 512 D\textsc{imension} 8 H\textsc{eads} S-\textsc{vector} M\textsc{odel} T\textsc{rained with} V\textsc{oxCeleb-1+2}}
 \label{tab:gg}
 \centering
\begin{tabular}{c|c|c|c}
\hline
Method & EER & DCF (0.01) &DCF (0.001) \\ \hline
\hline
\textbf{6L-512D-8H-S-300} &\textbf{2.67} &\textbf{0.30} &\textbf{0.44}\\
6L-512D-8H-S-100 &2.96 &0.35 &0.53\\
6L-512D-8H-S-500 &2.63 &0.28 &0.49\\ \hline
\multicolumn{4}{p{7.0cm}}{Note: 6L-512D-8H-S-300 means it is a 6 encoder layer (L), 512 Adim (D), and 8 head (H) s-vector (S) model with chunk length of 300.}
\end{tabular}
\end{table}

\subsection{Effect of Different Tapping Positions}
\begin{figure}[h]
  \centering
  \includegraphics[width=0.75\linewidth]{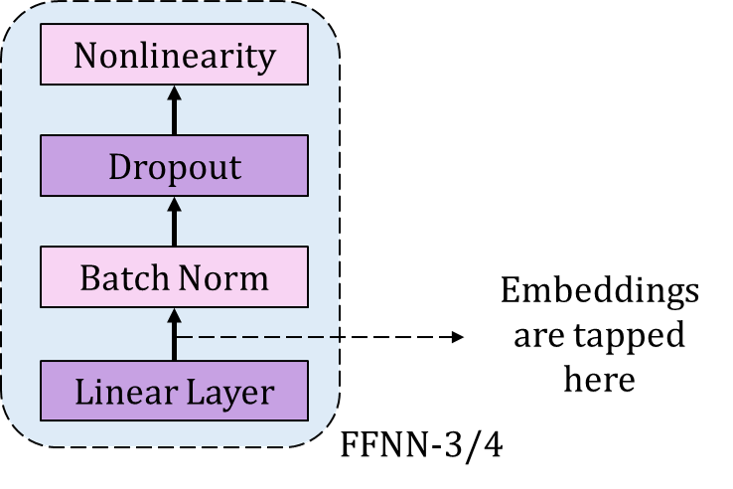}
  \caption{Embedding tapping location in an FFNN}
  \label{fig:tap}
\end{figure}

We analyzed the effect of deriving embeddings from the affine part of FFNN-4 instead of FFNN-3. \% EER and DCF for both of these positions of a 6L-512D-8H-S model trained on VoxCeleb-1+2 are presented in Table~\ref{tab:d}. 6L-512D-8H-S-F3 means that the embeddings are tapped from the affine part of FFNN-3 of the 6L-512D-8H-S model. We see that FFNN-3 gives better embeddings than FFNN-4. Fig.~\ref{fig:tap} shows the tapping of embeddings from the affine part of FFNN-3 and FFNN-4.

\begin{table}[h]
\renewcommand{\arraystretch}{1.3}
\caption{E\textsc{ffect of} T\textsc{apping} L\textsc{ayer on} 6L-512D-8H-S M\textsc{odel}}
 \label{tab:d}
 \centering
\begin{tabular}{c|c|c|c}
\hline
Layer & EER & DCF (0.01) &DCF (0.001) \\ \hline
\hline
\textbf{6L-512D-8H-S-F3} &\textbf{2.67} &\textbf{0.30} &\textbf{0.44}\\
6L-512D-8H-S-F4 &3.12 &0.33 &0.54\\ \hline
\multicolumn{4}{p{7.0cm}}{Note: 6L-512D-8H-S-F3 means it is a 6 encoder layer (L), 512 Adim (D), and 8 head (H) s-vector (S) model with embeddings extracted from FFNN-3 (F3).}
\end{tabular}
\end{table}

\subsection{S-vector and X-vector Ensemble}
\begin{figure}
  \centering
  \includegraphics[width=0.8\linewidth]{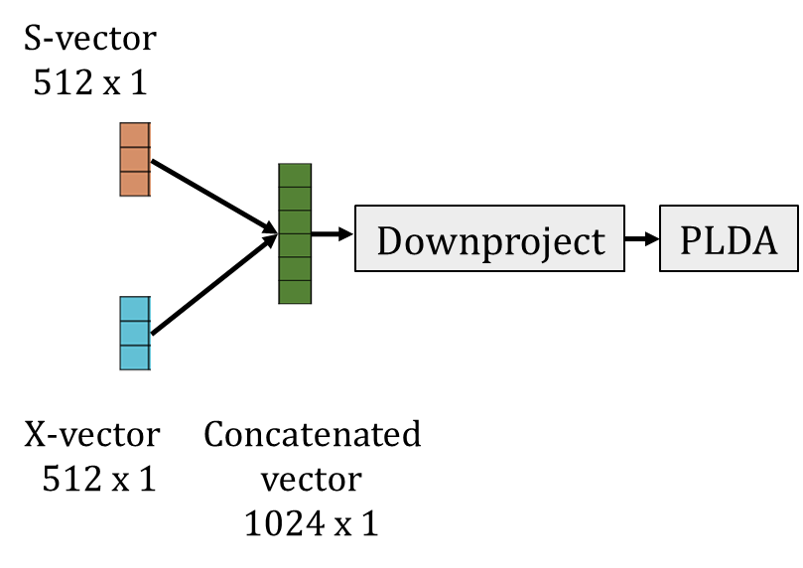}
  \caption{S-vector and x-vector ensemble by concatenation}
  \label{fig:concat}
\end{figure}

As the s-vector and x-vector models are completely different in their fundamental architecture, we expected that the ensemble of these two models might improve the results. Therefore, the s-vectors of 6L-512D-8H-S and x-vectors from the baseline model were concatenated to obtain 1024-dimensional ensemble vectors. These vectors were then downprojected to 300 dimensions and then presented to PLDA for scoring, as shown in Fig.~\ref{fig:concat}. As expected, the ensemble gives a significant improvement, as shown in Table~\ref{tab:ensem}.

\begin{table}
\renewcommand{\arraystretch}{1.3}
\caption{P\textsc{erformance of} S\textsc{-vector} \textsc{and} X\textsc{-vector} E\textsc{nsemble}}
 \label{tab:ensem}
 \centering
\begin{tabular}{c|c|c|c}
\hline
Method & EER & DCF (0.01) &DCF (0.001) \\ \hline
\hline
x-vector baseline &3.05 &0.33 &0.5\\
6L-512D-8H-S &2.67 &0.30 &0.44\\
\textbf{Ensemble} &\textbf{2.35} &\textbf{0.26} &\textbf{0.36}\\
\hline
\end{tabular}
\end{table}

 \section{Proposed Transformer Encoder Speaker Authenticator (TESA)}
 In PLDA, given a pair of enrollment and test embeddings, a log-likelihood score is computed to verify any identity claim. We propose a new speaker verification architecture, as a replacement for PLDA, in this section. This architecture is inspired by BERT.

\subsection{TESA Architecture}
BERT's next sentence prediction task is the inspiration for the Transformer encoder speaker authenticator (TESA). As mentioned before, the next sentence prediction task predicts whether the second text sentence in the sentence pair follows the first sentence. Therefore, we hypothesized that an architecture similar to BERT can be trained with embeddings from pairs of speech utterances to predict whether it is from the same speaker. In this case, the input is s-vector embeddings.

The data preparation for TESA is similar to the data preparation for BERT. However, we do not perform any masking because we are only interested in finding whether both utterances belong to the same speaker. It should also be noted that TESA operates on embeddings corresponding to a pair of spoken utterances and BERT operates on pairs of text sentences.

To train TESA, CMN was performed, and nonspeech frames were removed from every utterance. Then, s-vectors were extracted for every nonoverlapping 300 frame chunk of an utterance. Therefore, each utterance results in a set of embeddings, and every trial pair results in two sets of embeddings: Utterance-$1$ s-vectors of dimension \textit{512} $\times $ \textit{L} and Utterance-$2$ s-vectors with dimension \textit{512} $ \times $ \textit{M}, where $L$ and $M$ are the number of chunks in the respective utterances.

\begin{figure}
  \centering
  \includegraphics[width=0.8\linewidth]{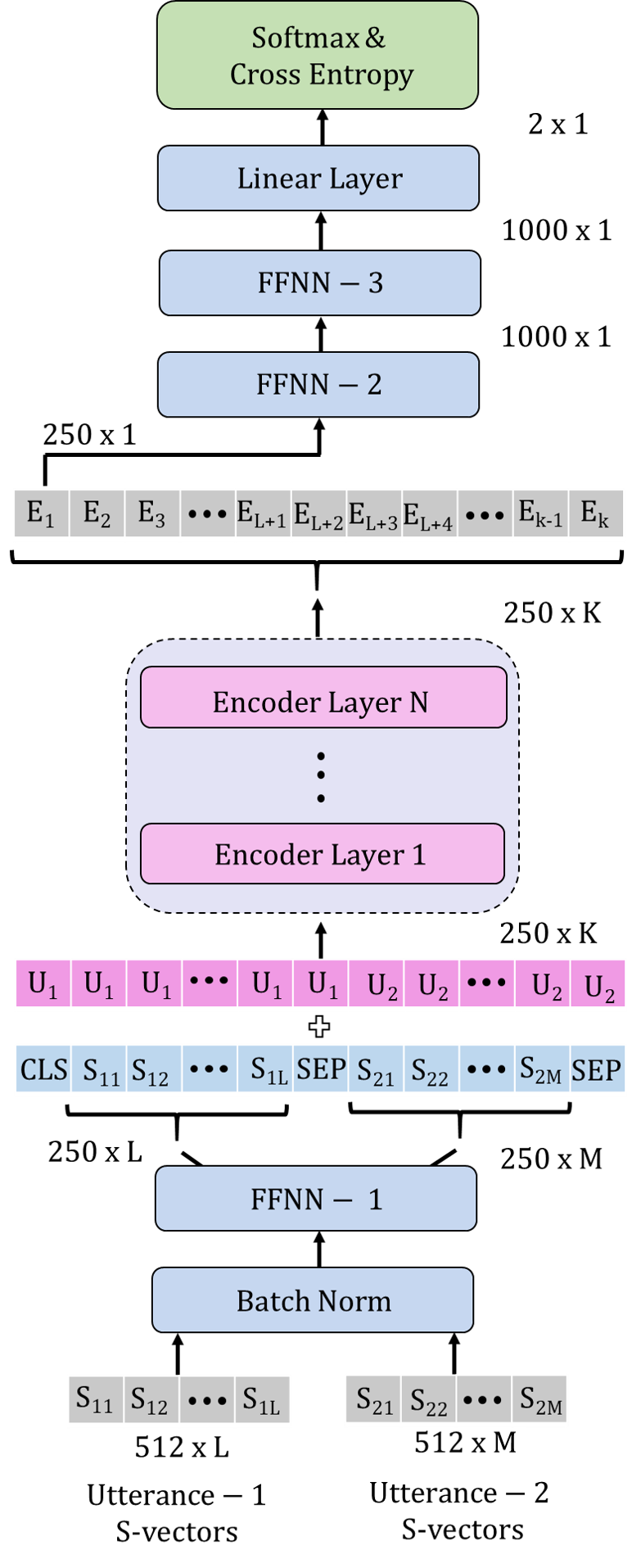}
\caption{Transformer Encoder Speaker Authenticator (TESA)}
  \label{fig:tesa}
\end{figure}

Similar to BERT's data preparation, classifier (CLS) and separator (SEP) tokens are taken, as shown in Fig~\ref{fig:tesa}. Unlike BERT, SEP demarcates the boundary between the two utterances instead of sentences. The resultant has the dimension \textit{512} $ \times $ \textit{K}, where $K=L+M+3$.

The s-vectors are then added with learnable utterance embeddings (U$_1$, U$_2$) to indicate that the two sets of s-vectors belong to two different utterances. This is similar to adding sentence embeddings in BERT. We do not add position embeddings to TESA because the prediction should not depend on the order of the s-vector embeddings. The resultant is then fed to FFNN-1, which takes it to \textit{250} $\times$  \textit{K}. This is now fed to the Transformer encoder.

At the end of the Transformer encoder layers, $E_1$ corresponding to the input CLS token is tapped and fed to the next layers for classification, as shown in Fig.~\ref{fig:tesa}. Other outputs of the encoder ($E_2, E_3,..., E_K$) are not considered. FFNN-2 takes the \textit{250} $ \times $ \textit{1} vector to the \textit{1,000} $ \times $ \textit{1} vector, and then FFNN-3 takes this \textit{1,000} $ \times $ \textit{1} vector to another \textit{1,000} $ \times $ \textit{1} vector. The linear layer projects the resultant vector to \textit{2} $ \times $ \textit{1}. Then, softmax is applied, and the model is trained on binary cross-entropy loss to predict whether the embeddings belong to the same speaker. We obtained the score for a given trial pair by subtracting the logits corresponding to the different-speaker classes from the same speaker class. All the nonlinearities used are ReLUs. We have performed batch normalization instead of layer normalization, similar to the s-vectors architecture.
\begin{figure}
  \centering
  \includegraphics[width=1.0
  \linewidth]{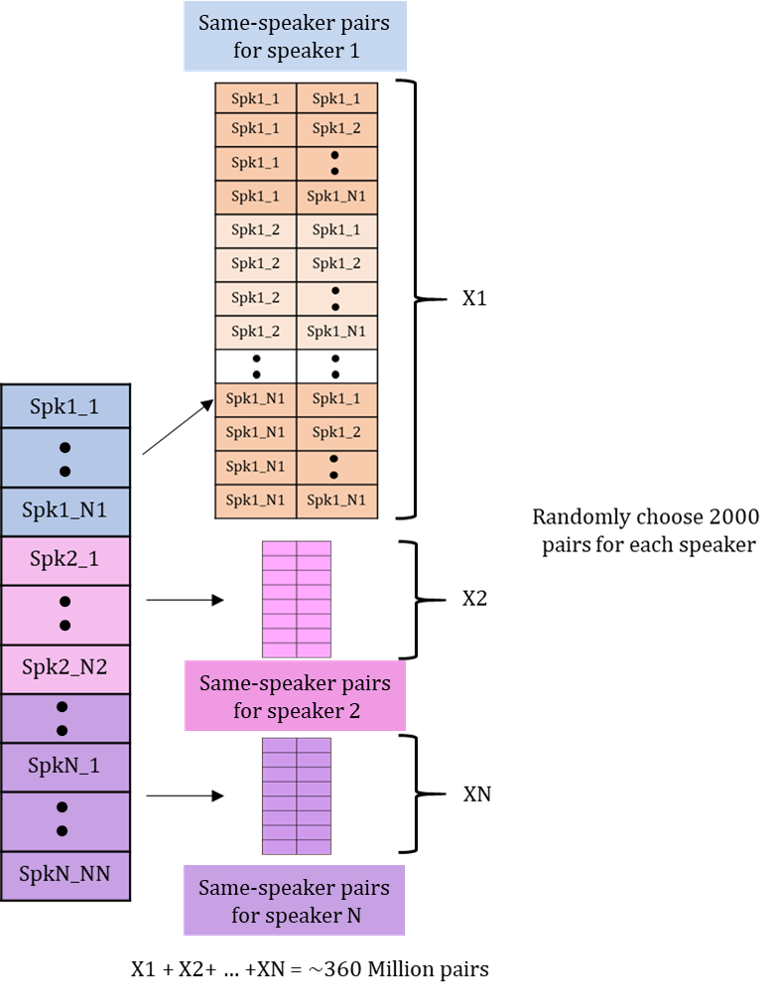}
  \caption{Same-speaker dataset creation}
  \label{fig:same}
\end{figure}
\begin{figure}
  \centering
  \includegraphics[width=1.0\linewidth]{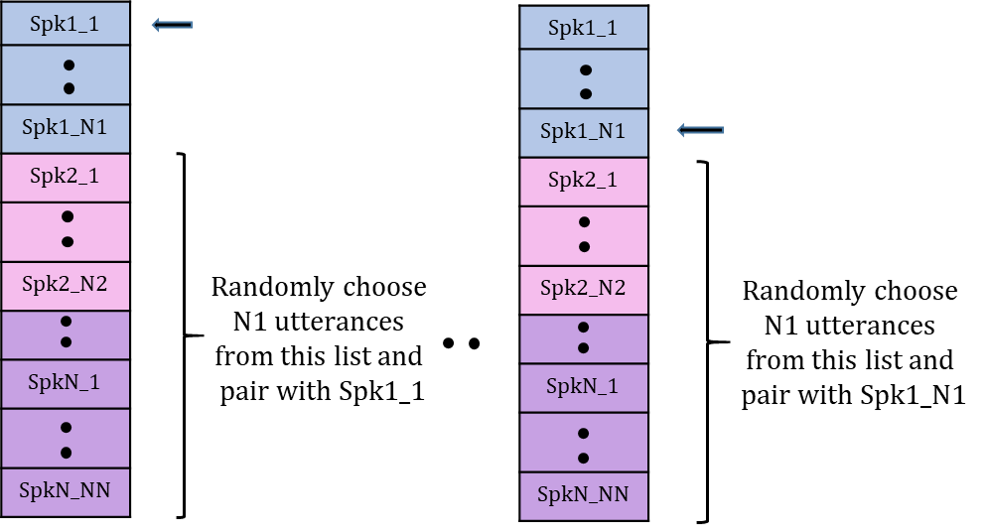}
  \caption{Different-speaker dataset creation for one speaker}
  \label{fig:different}
\end{figure}
\subsection{Training Details}
We used the same data PLDA uses for baseline x-vectors and extracted the s-vectors for chunks, as discussed in the previous subsection. The data have approximately 1.2 million utterances. After performing CMN and nonspeech frame removal, we created two sets of paired data: same-speaker and different-speaker. To create the same-speaker dataset, every utterance in the original dataset was paired with all possible same-speaker utterances, as shown in Fig \ref{fig:same}. This resulted in approximately 360 million pairs. To obtain 360 million pairs for the different-speaker dataset, every utterance of a particular speaker was paired with X randomly chosen different-speaker utterances, where X is the number of same-speaker utterances available, as shown in Fig \ref{fig:different}. As the number of pairs was too high for the experiment, we randomly chose 2,000 pairs per speaker from both sets. This ensured the balance between the two classes. Some speakers resulted in fewer than 2,000 pairs due to fewer utterances pertaining to that speaker in the VoxCeleb dataset. For such speakers, all the available pairings were taken. These two datasets were then combined and shuffled to create the training dataset for TESA. All the experiments for TESA were performed in ESPnet. The hyperparameters are presented in Table~\ref{tab:tesv}.

\begin{table}
\renewcommand{\arraystretch}{1.3}
\caption{T\textsc{raining} D\textsc{etails of} TESA}
 \label{tab:tesv}
 \centering
\begin{tabular}{c|c}
\hline
Hyperparameters &Details\\ \hline
\hline
Encoder Layers &9\\
Position Encoding  &No\\
Encoder Units      &1,024\\
Normalize Before   &True \\
Learning Rate      &10  \\ 
Batch Size         &2,000 \\
Optimizer          &Noam\\
Dropout             &0.1 \\
Gradient-clip       &5\\
Warm-up Steps      &25,000\\ 
Normalization      &Batch Norm\\
Adim &250\\\hline
\end{tabular}
\end{table}

\section{Results with the Proposed TESA and Discussion}
\begin{figure}
  \centering
  \includegraphics[width=0.9\linewidth]{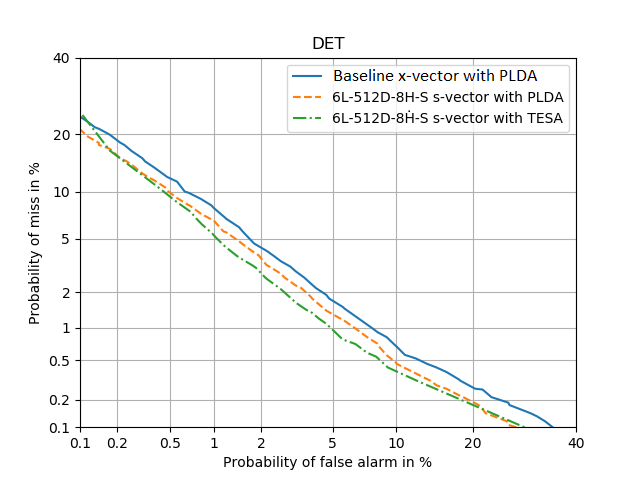}
  \caption{Comparison of DET curves of s-vectors with TESA, s-vectors with PLDA and the baseline x-vector with PLDA}
  \label{fig:DET_tesa}
\end{figure}

All the results in section IV were obtained by PLDA scoring. In this section, we analyze the results of the Transformer encoder speaker authenticator (TESA) proposed in this paper. Table~\ref{tab:tesa} shows the results obtained with the 9-layer TESA model. The number of parameters in TESA is 8.3 million. In Table~\ref{tab:tesa}, it can be seen that TESA outperforms the conventional PLDA by a significant margin. We obtain a relative improvement of 11\% in EER, and the DCF values are also better than those of conventional PLDA. DET curves for s-vectors with TESA, s-vectors with conventional PLDA and the baseline x-vectors with conventional PLDA are shown in Fig.~\ref{fig:DET_tesa}.

\begin{table}
\renewcommand{\arraystretch}{1.3}
\caption{C\textsc{omparison of} TESA \textsc{and} PLDA \textsc{for} S\textsc{-vectors from} 6L-512D-8H-S}
 \label{tab:tesa}
 \centering
\begin{tabular}{c|c|c|c}
\hline
Encoder Layers & EER &DCF (0.01) &DCF (0.001) \\ \hline
\hline
s-vectors + PLDA &2.67 &0.30 &0.44\\
\textbf{s-vectors + TESA} &\textbf{2.37} &\textbf{0.28} &\textbf{0.39}\\\hdashline
\end{tabular}
\end{table}

\section{Conclusions and Future Work}
In this work, we proposed deriving speaker embeddings from a speaker classification architecture based on Transformer's encoder. We call these embeddings s-vectors. The s-vectors obtained from our model trained on VoxCeleb-1+2 datasets outperformed the \% EER of the baseline x-vector system trained on the same data, and gave a relative improvement of $12.5\%$. When trained only on the smaller VoxCeleb-1 dataset, the s-vectors again outperformed the \% EER of the baseline x-vector system trained on the same data, and resulted in a relative improvement of $10\%$. Our model is also better in terms of DCF. In addition to s-vectors, we also proposed replacing conventional PLDA-based speaker verification with a new architecture named the Transformer encoder speaker authenticator (TESA). TESA outperformed the \%EER of PLDA trained on the same dataset by 11\% relative and has better DCF values. TESA with s-vectors jointly gives a relative improvement of 22.3\% over the baseline x-vectors with PLDA on VoxCeleb-1 trials when trained on VoxCeleb-1+2 data. In the future, we would like to explore different pretraining methods and loss functions to improve the performance of the proposed architectures. We would also like to explore the capabilities of TESA trained in an end-to-end manner by feeding utterance pairs directly for speaker verification.

\bibliographystyle{IEEEtran}

\bibliography{s-vectors-2}

\end{document}